\documentclass[a4paper]{jpconf}
\usepackage{graphicx}
\begin{document}
\title{Jarzynski Equality for an Energy-Controlled System (Proceedings of nanoPHYS' 11)}

\author{Hitoshi Katsuda$^1$ and Masayuki Ohzeki$^2$}

\address{$^1$ Department of Physics, Tokyo Institute of Technology, Oh-okayama, Meguro-ku, Tokyo 152-8551, Japan}
\address{$^2$ Department of System Science, Graduate School of Informatics, Kyoto University, Yoshida-Honmachi, Sakyo-ku, Kyoto 606-8501, Japan}

\ead{$^1$ katsuda@stat.phys.titech.ac.jp}

\begin{abstract}
We extend the Jarzynski equality, which is an exact identity between the equilibrium and nonequilibrium averages, to be useful to compute the value of the entropy difference by changing the Hamiltonian.
To derive our result, we introduce artificial dynamics where the instantaneous value of the energy can be arbitrarily controlled during a nonequilibrium process.
We establish an exact identity on such a process corresponding to the so-called Jarzynski equality. 
It is suggested that our formulation is valuable in a practical application as in optimization problems.
\end{abstract}

\section{Introduction}
To decrease the power loss, we often design the shortest paths to connect with each components in electric integrated circuits.
Not only the electric loss but also any kinds of the cost are demanded to be minimized as far as possible in industrial products.
Such problems can be formulated into a more generic task to minimize or maximize a real single-valued function of multivariables, which is called optimization problems \cite{np,Opt}.
Several solvers for optimization problems have been invented in the context of the dynamical process in statistical physics \cite{SA,QA}.
In these methods, the system is driven to be trapped at the global minimum of the complicated valley structure.

In the present study, we approach this issue in a non-standard way with the recent progress in statistical mechanics.
We propose a method to measure the difference between the approximate result given by solvers of optimization problems and the true answer.
In practice, we do not always know the ground state energy, however the minimum value of the entropy is known trivially.
Therefore the entropy can be an indicator for the deviation of the approximate solution from the accurate answer.

To goal of our study, we extend the identity proposed by Adib, which is useful to estimate the entropy difference in changing the Hamiltonian in artificial dynamics \cite{isoje}.
This identity is inspired by the Jarzynski equality, which plays a key role to connect equilibrium states at beginning and end with a nonequilibrium process \cite{originalje}.
However the original formulation given by Adib is considered only for the isoenergy process, on which the energy is fixed to be a constant value, while the Hamiltonian changing.
We extend the identity of the entropy difference in the isoenergy process to an energy-controlled process \cite{mine}.
\section{Formulation}
In order to arbitrarily control the energy, we introduce an artificial field $\mathbf{F}=(\mathbf{F}_x, \mathbf{F}_p)$ to the Hamilton dynamics. 
The equations of the modified dynamics are
\begin{equation}
\dot{\mathbf{x}}=\frac{\partial H}{\partial \mathbf{p}}+ \mathbf{F}_x(\mathbf{\Gamma}),\>\>\>
\dot{\mathbf{p}}=-\frac{\partial H}{\partial \mathbf{x}}+ \mathbf{F}_p(\mathbf{\Gamma}).
\label{eq:Hamilton dynamics}
\end{equation}
The above $\mathbf{\Gamma}=(\mathbf{x},\mathbf{p})$ describes a point on the phase space.
The energy follows an arbitrary function of time $E(t)$, while the Hamiltonian changing from $t=0$ to $\tau$, if we choose the functional form of $\mathbf{F}=(\mathbf{F}_x, \mathbf{F}_p)$ as
\begin{equation}
\mathbf{F}(\mathbf{\Gamma}) = \frac{\mathbf{X}}{\mathbf{X}\cdot \partial _{\mathbf{\Gamma}}H}\biggl( \frac{\rm d \it E}{\rm d \it t}-\frac{\partial H}{\partial t}   \biggr).
\label{eq:energy reservoir}
\end{equation}
Here $\mathbf{X}$ is an arbitrary vector on the phase space satisfying $\mathbf{X}\cdot \partial _{\mathbf{\Gamma}}H\neq 0$.

>From equations (\ref{eq:Hamilton dynamics}) and (\ref{eq:energy reservoir}), we can easily confirm that the energy is accurately controlled as $E(t)$ since we have
\begin{equation}
\frac{\rm d \it H}{\rm d \it t}=\frac{\partial H}{\partial \mathbf{\Gamma}}\cdot \dot{\mathbf{\Gamma}}+\frac{\partial H}{\partial t}=\frac{\rm d \it E}{\rm d \it t}.
\label{eq:control}
\end{equation}

Under the special dynamics (\ref{eq:Hamilton dynamics}), the ensemble density $\rho _t(\mathbf{\Gamma})$ evolves following the Liouville equation:
\begin{equation}
\rho _t(\mathbf{\Gamma }_t) = \rho _0(\mathbf{\Gamma}_0) \rm e^{\it -t \overline{\rm \Lambda _{\it t}}(\mathbf{\Gamma}_t)},
\label{eq:Liouville eq}
\end{equation}
where
\begin{equation}
\overline{\Lambda _t}(\mathbf{\Gamma}_t) = \frac{1}{t}\int _0 ^t \rm d \it t' \rm \Lambda \it (\mathbf{\Gamma }_{t'}).
\label{eq:Lambda bar}
\end{equation}
Equation (\ref{eq:Lambda bar}) is the time average of the ``phase space compression factor" $\Lambda (\mathbf{\Gamma}_t) = \partial _{\mathbf{\Gamma}_t} \cdot \dot{\mathbf{\Gamma}_t}$ along the trajectory that connects $\mathbf{\Gamma}_0$ to $\mathbf{\Gamma}_t$ \cite{liouville}.
This factor appearing in the dynamics of the ensemble density $\rho _t(\mathbf{\Gamma}_t)$ will play the central role to estimate the entropy as shown below.
Similarly to the original Jarzynski equality, the system is assumed to be in an equilibrium state at the initial time $t=0$. 
The distribution  $\rho _0(\mathbf{\Gamma}_0) $ at the initial time is set to be the microcanonical distribution at $E = E(0)$:
\begin{equation}
\rho _0(\mathbf{\Gamma}_0) =\frac{\delta (H(\mathbf{\Gamma}_0)-E(0))}{\Omega _0},
\label{eq:inital}
\end{equation}
where $\Omega _0$ is the number of states at $t=0$. 

Let us consider the average of $\rm e^{\tau\overline{\Lambda _{\tau}}}$ over all possible realizations from $t=0$ to $\tau$:
\begin{equation}
\langle \rm e^{\it \tau\overline{\Lambda _{\tau}}} \rangle   = \int \rm d \it \mathbf{\Gamma}_{\rm \tau}\rho _{\tau}(\mathbf{\Gamma}_{\tau}) \rm e^{\tau \overline{\Lambda _{\tau}}} \\
   = \int \rm d  \mathbf{\Gamma}_{\tau} \frac{\delta (\it H(\mathbf{\Gamma} _{\rm 0}\it)-E(\rm 0))}{\Omega _0}. 
\label{average}
\end{equation}
In the second transformation, equations (\ref{eq:Liouville eq}) and (\ref{eq:inital}) have been used.
Note that we control the energy of the system as $H(\mathbf{\Gamma}_t) = E(t)$, and we thus find
\begin{equation}
H(\mathbf{\Gamma}_t) -E(t)=H(\mathbf{\Gamma}_0) -E(0)
\label{eq:control}
\end{equation}
for any $t \in [0, \tau]$.
Therefore $H(\mathbf{\Gamma} _0)-E(0)$ in the Dirac delta function of equation (\ref{average}) can be replaced by $H(\mathbf{\Gamma} _{\tau})-E(\tau)$. 
Equation (\ref{average}) can be reduced to
\begin{equation}
\langle \rm e^{ \tau\overline{\Lambda _{\tau}}} \rangle = \rm e^{ \Delta \it S}.
\label{eq:myje}
\end{equation}
Above $ \Delta S = \ln (\Omega _{\tau}/\Omega _{0})$ is the entropy difference between the equilibrium states at the different energy values $E(0)$ and $E(\tau)$. 
Therefore we can obtain the entropy difference after the nonequilibrium procedure following the schedule of the energy $E(t)$.
If we estimate the entropy difference with such a naive method as directly calculating the entropy at the initial and final times, we have to investigate the entropy twice. 
On the other hand, our formula (\ref{eq:myje}) gives the entropy difference by taking just a single average of $\rm e^{\tau \overline{\Lambda _{\tau}}}$.
This means that we can examine the entropy difference by the direct use of the resulting ensemble after the nonequilibrium process.

\section{Application to optimization problems}
We here show that our equality can be used for a quantitative estimation, which indicates how much an approximate solution differs from the true solution.
Let us consider an arbitrary potential energy $V(\mathbf{x})$ with continuous variables $\mathbf{x}$ which has no degeneracy at the ground state, and assume that a solution with the energy $\mathcal{E}$ has been obtained. 
We consider the following Hamiltonian 
\begin{equation}
H=\mathbf{p}^2+\frac{t}{\tau}V(\mathbf{x}),\label{eq:Hforopti}
\end{equation}
At the first stage of the dynamics, the system can visit all the locations since the potential energy is small.
By increasing the coefficient of the potential energy, the particle can recognize the energy barriers and be trapped into local minima.
Let us consider the schedule of the energy from $E(0) = \mathcal{E}'$ to $E(\tau) =\mathcal{E}$, where $\mathcal{E}'>\mathcal{E}$. 
The entropy at the initial time $S(0)$ can be calculated easily since the system equals to non-interacting particles with the mass $m=1/2$ at $t=0$.
On the other hand, the entropy at the final time $S(\tau) = S(0)+\Delta S$ can be estimated by calculating $\Delta S$ from our identity.

The detailed protocol is described below.
First, we randomly choose an initial condition $( \mathbf{x}_0, \mathbf{p}_0)$ from the set $\{ (\mathbf{x}$, $\mathbf{p}) | H(\mathbf{x}, \mathbf{p}; t=0) = \mathcal{E}' \}$.
Note that the initial Hamiltonian depends only on $\mathbf{p}$. 
We obtain the path toward a phase point with the lower energy following the energy-controlled dynamics (\ref{eq:Hamilton dynamics}) under the given initial condition. 
Some initial conditions result in the divergence of the factor appearing on the right-hand side of equation (\ref{eq:energy reservoir}) when the system is trapped in a local minimum with $\partial_{\mathbf{\Gamma}_t}H=0$. 
Such a divergence means that the energy can not decrease toward $\mathcal{E}$ and the system evolving from such initial conditions are unable to reach any points on the phase space at $t=\tau$.
Therefore such samples should be excluded in taking the average of $\exp\tau \overline{\Lambda _{\tau}}$.
Finally, we take the average of $\exp\tau \overline{ \Lambda _{\tau}}$ only in the case of the absence of such divergences, and we obtain $\Delta S$ appearing on the right-hand side of our equality (\ref{eq:myje}).

In the case of optimization problems, the value of the energy at the ground state cannot be known in advance.
On the other hand, the minimum value of the entropy is trivial as $-\infty$ in the case of classical systems.
Therefore $S(\tau) \to -\infty$ implies that the obtained solutions after the nonequilibrium process for the given potential energy are close to the minimum point.
In other words, estimating the difference of the entropy can be an indicator how much the approximate solution differs from the actual minimum.

The detailed estimation by use of a simple instance is given in the reference \cite{mine}.
Let us here emphasize another advantage in use of the entropy difference to measure the deviation from the true solution.
As described in figure 1, if $\Delta E =\mathcal{E} _2-\mathcal{E} _1$ takes a common value in both case, the entropy gap $\Delta S$ can become quite different, since the entropy rapidly decreases when the system breaks through local minima of the potential energy. 
The energy cannot lead us to any information on the difference coming from the energy structure. 
Therefore the entropy difference can be a good measure of the deviation from the true solution in optimization problems.

\begin{figure}[h]
\includegraphics[scale=0.4]{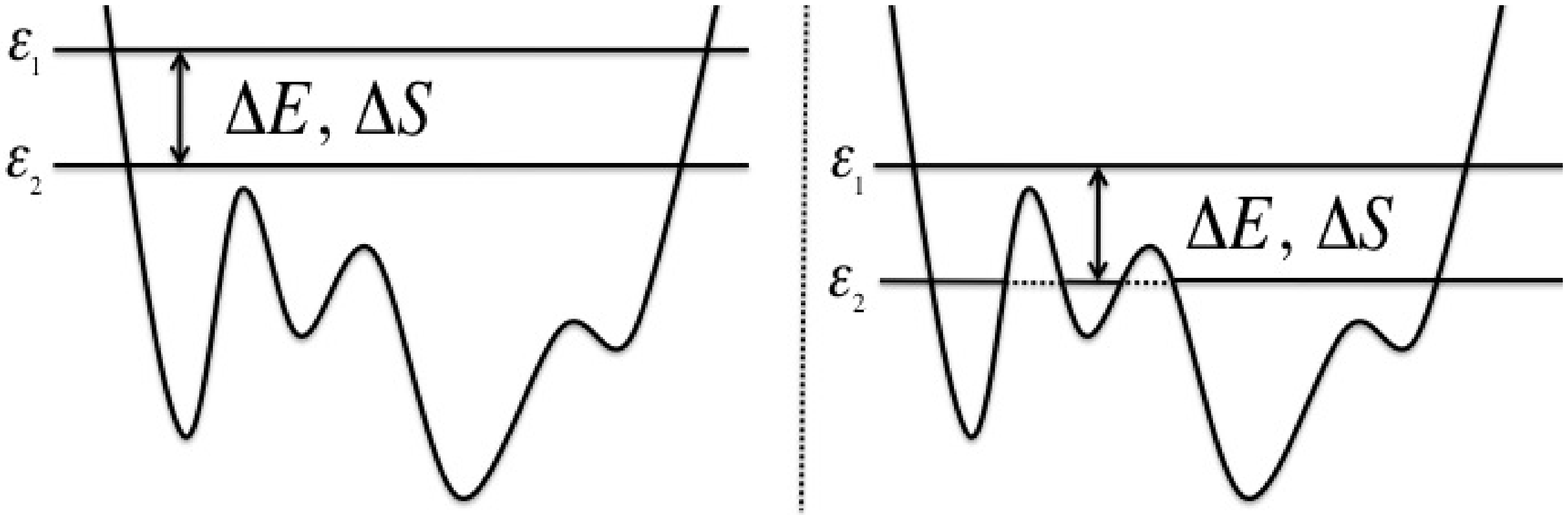}
\caption{Case that equal $\Delta E$ but different $\Delta S$.}
\label{f2}
\end{figure}

\section{Conclusion}
We have extended the Adib identity inspired by the Jarzynski equality to an energy-controlled system, and shown the possibility as the measure of the deviation of the approximate solution from the true answer.
However our study is still short of consideration, since the efficient dynamics to search for the ground state have not considered.
In addition we have trouble in numerical computation, since the value of the entropy of an approximate solution is compared with an infinite value in our protocol.
To solve these problems, we might establish the quantum version of our identity. 
In the case of a quantum system, the entropy at the ground state is equal to $0$ which is clearly more suitable for the quantitative estimation than $- \infty$ of classical systems. 
Furthermore, the tunneling effect can be useful, as quantum annealing, for searching for the global minimum of the complicated energy landscape as in spin glasses, with which most of the optimization problems are closely related \cite{sg}.

We also remark that the quantum version of our formulation can also open the possibility of quantum computation.
By use of the quantum degrees of freedom, we can implement our identity in the quantum computation \cite{annealing}. 
The same techniques as in the literature would be available to suggest the algorithm to estimate the entropy as well as the free energy.

\end{document}